\documentclass[aps,prl,twocolumn,showpacs,preprintnumbers]{revtex4}
\usepackage{graphicx}
\begin{document}
\newcommand{\vev}[1]{\langle{#1}\rangle}

\preprint{\sf Version 1 (\today)}
\title{A New Class of Solutions to 
the Strong CP Problem with a Small Two-Loop $\theta$}
\author{Darwin Chang}
\affiliation{Physics Department, National Tsing-Hua University,
Hsinchu 300, Taiwan}

\author{Wai-Yee Keung}
\affiliation{Department of physics, University of Illinois at Chicago,
Illinois 60607-7059, USA}

\date{\today}

\begin{abstract}
We present a new class of models which produce zero $\theta_{\rm
QCD}$ angle at the tree and one-loop level due to hermiticity of
sub-blocks in the extended quark mass matrices.  The structure can
be maintained typically by non-abelian generation symmetry.  Two
examples are given for this class of solutions.
\end{abstract}

\maketitle

\vskip.1cm \noindent{\bf Introduction} \quad
While the Standard Model (SM) has been enjoying fantastic success,
it does have many loose ends which are potentially our guidepost to
the new physics of the future.  Two of the most significant loose
ends are strong CP problem and the fermion mass hierarchy.  Within
the SM, the Yukawa couplings give rise to the fermion masses of
all three generations and their mixings including the CP
violation.  Indeed it was first observed by Kobayashi and
Maskawa\cite{Kobayashi:fv} (KM) that only  two generations cannot
support any CP violating phase. The fact that all three generations
have to be involved to create a CP violating phenomena, makes
KM model an extremely subtle and beautiful model for CP violation.
It also makes CP violation tightly connected with flavor physics.

One of the weaknesses of the SM is that it does not address the
issue of why fermion masses are so awkwardly different.  The top
quark mass is five orders of magnitude bigger than electron mass
which is in turn bigger than the neutrino mass by another four to
five orders.  This is the fundamental issue in flavor physics.
However, there is an even more serious problem if one looks into
the CP issue in the KM mechanism.  The mechanism allows a tree
level CP violating parameters $\theta$ associated with strong
interaction. The experimental limit on the neutron electric dipole
moment (EDM) requires this parameters to be of order $10^{-10}$ or
smaller. The KM mechanism does not address why $\theta$ is so
small.  There are two levels to this problem, the tree and the
loop levels.  Since CP violating phase in KM is a part of the
dimension-four Yukawa couplings, just like the tree level $\theta$
parameter, there is no reason why $\theta$ can not be a bare
parameter of the model with a natural value of order one. However,
it should be noted that, by being tightly connected with the
flavor physics as described above, the KM mechanism has already
embedded in itself a natural mechanism to suppress the loop
correction to $\theta$.  It has been shown that if one
heuristically set $\theta$ to zero at the tree level, the loop
correction will not occur till  three-loop level (with two weak
and one strong loop)\cite{Shabalin:zz}, and logarithmic divergent
correction to $\theta$ will appear only at the 14th order of  the
electroweak coupling $g_2$ (or, at the 7-loop
level)\cite{Ellis:1978hq,Khriplovich:1993pf}. 
Even if one put in the Planck scale as the estimate for the cut-off, the
divergent correction produces only a minute value for $\theta$ just like
the 3-loop finite corrections.

This nice loop property is a direct result of the coupling between
flavor physics and CP in the KM model.  It indicates that the
strong CP problem is in a sense only a tree level problem.   All
we need is to look for a mechanism extending  SM to suppress the
tree level $\theta$ parameter.  After that, the loop correction
will take care of itself.   Such mechanism does not have to be a
low energy phenomenon.  It can be some features embedded in a high
energy theory such as GUT or string theory.  For example, a
popular class of models of this type is the Nelson-Barr
mechanism\cite{Nelson:hg,Barr:qx} in which a (softly broken or gauged)
flavor symmetry and spontaneously broken CP symmetry are used at
high energy in a GUT-like theory to suppress the tree level
$\theta$.  The phase of the KM model is generated by introducing
additional heavy vectorial fermions which can have CP violating
mixing with the ordinary fermions.  In models such as this, one
typically have a one-loop induced $\theta$ at the higher energy
scale.  Such contributions are typically not suppressed by the
heavy scale and it is up to the adjustment of the model parameters
to make such contributions small enough.  While flavor symmetry
was used in examples provided by Ref.\cite{Nelson:hg,Barr:qx,Bento:ez}, it
can potentially be replaced by some other symmetry. Still another
recent example, Ref.\cite{Glashow:2001yz} involving even more
direct use of flavor symmetry, adopts both flavor and CP symmetry
to make the up (down) quark mass matrix lower (upper) triangular
with real diagonal elements.  This also guarantees that the tree
level $\theta$ is zero while the model still has enough parameters
to create the KM phase of any magnitude. Clearly such mechanism
can be easily embedded in GUT or SUSY context.

While not all the proposed solutions to the strong CP problem are
strongly associated with flavor physics (for example, the
Peccei-Quinn solution\cite{ref:PQ} can be quite independent of
flavor), it should be interesting to speculate that the solution
to the strong CP problem may be the side product of high energy
theory that addresses the issues in the flavor problem.

In this letter, we wish to propose a new class of solutions to the
strong CP problem that can also potentially address the flavor
physics issue.  The new class of models we propose involves
additional heavy fermions in high energy and a special mass matrix
pattern for this extended set of fermions such that it results in
vanishing strong CP $\theta$ even at one-loop level.  The small
two-loop $\theta$ in the high energy theory naturally produces an
effective KM model at low energy with small enough tree level
effective $\theta$ that predicts a small but potentially
measurable neutron EDM.  The special mass pattern can be the
result of some family symmetry which can serve to tie up the issue
of strong CP problem with that of the flavor problem in the
future.

We shall illustrate the basic structure that characterizes this
class of models and then provide two examples using two different
kind of flavor group to produce the desired extended fermion mass
pattern.  Since it is not our purpose here to try to pin-point a
particular realistic model at this point, we shall only present
the simplest examples of such solution to the strong CP problem
and leave the issues related to the fermion mass hierarchy to the
future.  This is reasonable because even if there is a common link
to both problems, the strong CP problem seems to be more severe
while the flavor problem seems to be more tedious and more tied to
details of  model building.
\vskip.2cm  \noindent{\bf Minimal Model of Hermitian Mass Matrix}\quad
The idea is simply to add  vector-like fermions with the
same charge as the up and down quarks so that the larger mass
matrices are hermitian at tree level using flavor symmetry and
spontaneously broken CP symmetry. This will guarantee that the
tree level $\theta$ at high energy is zero.  To make the radiative
correction to $\theta$ small enough, one arrange additional
symmetry to make one-loop contribution vanish in the high energy
theory.  This will result in a low energy effective theory whose
CP violation is is KM in nature while effective tree level
$\theta$ is naturally small.

Our simplest example uses a horizontal symmetry $SO(3)_\parallel$
to achieve real determinants of quark mass matrices at both the
tree and one-loop levels. 
In addition to the three existing  generations, there are three
new generations, which are individually $SU_L(2)$ singlets and
vector-like, nonetheless, the hypercharges are chosen in a way
that they have same electric charges as the known generations.
Therefore they are labelled by
\begin{equation}
U_{Li}, U_{Ri}, D_{Li}, D_{Ri} \ ,
\end{equation} in an analogous
fashion with the known quarks,
\begin{eqnarray}
q_{Li} \equiv \left(\begin{array}{c} u_{Li} \\
d_{Li}\end{array}\right),
      u_{Ri},d_{Ri} \ .
      \end{eqnarray}
Let us stress that  their  only difference is in $SU_L(2)$
assignments.
A horizontal flavor symmetry $SO(3)_\parallel$ transforms every Weyl
fermion multiplet above in the ${\bf 3}$ representation,
labelled by generation index $i=1,2,3$.
There are new horizontal neutral (inert to $SU(2)_L \times U(1)$) Higgs
bosons, {\it i.e.} one quintet (symmetric traceless rank-2 tensor)
CP-even  $\phi_S$
and one triplet (antisymmetrc rank-2 tensor) CP-odd $\phi_A$.
The Yukawa couplings are
\begin{eqnarray}
   \bar  d_R(\mu_d+g_{dS}\phi_S +ig_{dA}\phi_A) D_L \nonumber\\
 + \bar  u_R(\mu_u+g_{uS}\phi_S +ig_{uA}\phi_A) U_L
\nonumber\\
 + \bar  D_R(\mu_D+g_{DS}\phi_S +ig_{DA}\phi_A) D_L  \nonumber\\
 + \bar  U_R(\mu_U+g_{US}\phi_S +ig_{UA}\phi_A) U_L  \nonumber\\
+ (h_d \bar d_R+h'_d \bar D_R) H^\dagger q_L
+ (h_u \bar u_R+h'_u \bar U_R) \tilde H^\dagger  q_L + {\rm H.c.}
\end{eqnarray}
with the usual $SU_L(2)$ Higgs doublet $H$ which couples to
fermions flavor-blindly.  We denote  $ \tilde H \equiv i\tau_2 H^*
$.  CP symmetry is assumed to be a good symmetry before any
symmetry breaking.  Since $d_R$ and $D_R$ are identical triplet,
one indeed can define $D_R$ to be the one that couples with $q_L$
and there $h_d= 0$ by convention.

The horizontal symmetry is completely broken\cite{Li:1973mq} by
VEV's of $\phi_S$ and $\phi_A$, which  together with $\vev{H}$
give rise to the following $6\times 6$ down quark mass matrix term
\begin{equation} (\bar d_R \  \bar D_R ) (M_6)
\left(\begin{array}{c} d_L \\ D_L \end{array}\right) \ ,
\end{equation}
\begin{eqnarray} M_6= \left(\begin{array}{cc}
  {\bf 0} & \mu_d + g_{dS}\vev{\phi_S} +ig_{dA}\vev{\phi_A}\\
 h'_d\vev{H^\dagger}{\bf 1} & \mu_D + g_{DS}\vev{\phi_S} +ig_{DA}\vev{\phi_A}
  \end{array}\right)
\end{eqnarray}
and similarly for up quark mass matrix.  The $6\times 6$ matrix
has real determinant when couplings are real, as required by the
imposed CP symmetry.

Integrating out the exotic generation $D$, we have the effective
mass matrix for the three light generations like CKM phenomenology.
More explicitly, the effective masses of the known 3 generations can
be understood as the amplitude given by the diagram below. It is a
kind of see-saw mechanism
in the quark sector.  Similar diagrams occur for the up-type quarks.
\begin{figure}
\centering
\includegraphics[width=8cm]{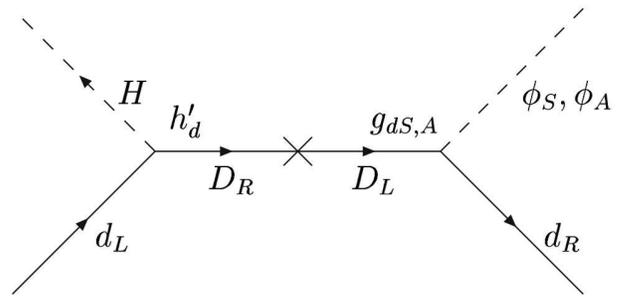}
\caption{A see-saw diagram for effective masses of the known 3 generations.
 }
\end{figure}

In the limit of $\vev{H}=0$, $d_L$ quarks  decouple and we have
the reduced mass matrix of  a size $6\times3$ instead,
\begin{eqnarray} (\bar d_R
\  \bar D_R ) \left(\begin{array}{c} M_d\\M_D
  \end{array}\right)
\left( D_L \right) \ ,
\end{eqnarray}
\begin{eqnarray} \qquad \left.\begin{array}{l}
    M_d=  \mu_d +   g_{dS}\vev{\phi_S} +ig_{dA}\vev{\phi_A} \ ,\\
    M_D=  \mu_D +   g_{DS}\vev{\phi_S} +ig_{DA}\vev{\phi_A} \ .
\end{array}\right.
\end{eqnarray}
First, we find a $6\times 6$ unitary matrix $V$ to transform the 
above $6\times3$ into one with zero entries in the upper $3\times
3$ block.
\begin{eqnarray}
V \left(\begin{array}{c}M_d
\\M_D\end{array}\right)=\left(\begin{array}{c} {\bf 0}
\\M'_D\end{array}\right)  \ ,
     \end{eqnarray}
by choosing the top three row vectors of $V$
perpendicular to the 3 column vectors in the mass matrix. This  is
possible because the 6 dimensional linear space is
larger than the 3 dimensional space spanned by the three column
vectors.
Furthermore, by using bi-unitary transformation,
we also rotate $D_L$  into $D'_L$
so that  $M'_D$ is diagonal. In this way, the massless states $d'_R$ and
the massive states $D'_R$ are generally  mixed  among the original $d_R$ and
$D_R$. Nonetheless, the three generations of $d_L$ remain massless and unmixed.

We include the effect $\vev{H}$ from the viewpoint of
perturbation. The mass terms involving $d_L$ are tabulated in the
matrix form,
\begin{eqnarray} (\bar d'_R \  \bar D'_R ) V \left(\begin{array}{c}
  {\bf 0}\\
 h'_d\vev{H^\dagger}{\bf 1}  \end{array}\right) d_L 
=(\bar d'_R \  \bar D'_R )
\left(\begin{array}{c}
  {\hat{m}}_d \\
  {\hat{m}'_d}
\end{array}\right) d_L     \ .
\end{eqnarray}
Including  $D'_L$, we have
\begin{eqnarray}
(\bar d'_R \  \bar D'_R ) \left(\begin{array}{cc}
  {\hat{m}}_d  & {\bf 0}\\
  {\hat{m}'_d} &  M'_D
\end{array}\right)
\left( \begin{array}{c} d_L\\ D'_L\end{array} \right)  \  .
\end{eqnarray}
As $M'_D\gg\hat{m}_d$, masses of  usual $d$-quarks
are basically given by diagonalization of the complex mass matrix
$\hat{m}_d$. Similar procedures also apply to the $u$-quarks.
Phenomenology of CKM mechanism follows.
\vskip.2cm \noindent{\bf Strong CP at the one-loop level}\quad
Since the hermiticity of each of the $3 \times 3$ block of the $M_6$
mass matrix is a consequence of the horizontal flavor and CP
symmetries, it is maintained even after $M_6$ receives radiative
corrections from the induced higher dimensional operators. 
To see this, start with a simple case by working out an effective
dim 5 operator of the form
\begin{equation}
  i\bar D_L (f_{SA} \phi_S\phi_A  +f_{AS}\phi_A\phi_S) d_R \ ,
  \end{equation}
which is induced at the loop level, like the figure below. 
\begin{figure}
\centering
\includegraphics[width=8cm]{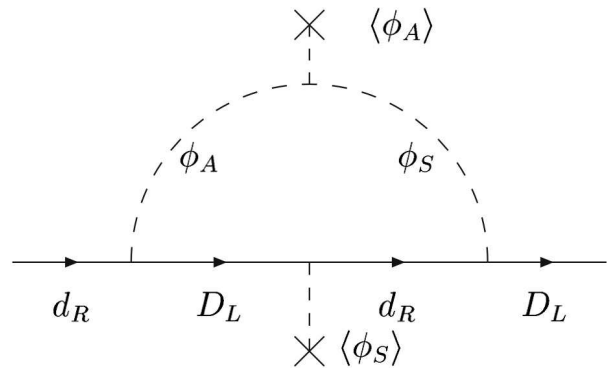}
\caption{\label{fig2} Induced mass matrix with block-hermitian property. }
\end{figure}
The hermitian conjugate of above operator has the form
$$  -i\bar d_R (f^*_{AS} \phi_S\phi_A  +f^*_{SA}\phi_A\phi_S) D_L \ . $$
However, there is a CP symmetry imposed in the beginning such that
the operator is invariant under
\begin{equation} \phi_S \to \phi_S \ ,\quad
  \phi_A \to -\phi_A \ ,\quad
\bar D_L (\cdots) d_R \to \bar d_R (\cdots ) D_L \ .
\end{equation}
This require $f_{SA}=f_{AS}^*$. This way, the
induced effective mass matrix
$$i(f_{SA} \vev{\phi_S}\vev{\phi_A}  +f_{AS}\vev{\phi_A}\vev{\phi_S})$$
remains hermitian. Generalization beyond this simple case is
straightforward.

With loop effects, the left upper and lower blocks begin
to deviate from the ones proportional to a unity matrix.  As the
texture at the tree level disappears, the determinant is no longer
real at the one-loop level, therefore there will be one-loop
contribution to $\theta$.  Although this model neither falls into
the Nelson--Barr class\cite{Nelson:hg,Barr:qx} nor the Glashow type of
models\cite{Glashow:2001yz}, they all suffer from the one-loop
contribution to the strong CP problem, which in principle can be
suppressed by some choices of hierarchy structure in parameters of
the models through fine-tuning. However, in our model, there is an
alternative and natural mechanism to remove the one-loop strong
$\theta$.

If we impose a discrete symmetry under which $u, d$ and all
$\phi_{A,S}$, $H$ fields are odd, while $U$ and $D$ are even, we have
$$ h_d=0\ , \quad g_{DS}=0 \ ,\quad  g_{DA}=0 \ . $$
We keep the term $\mu_d$ which only breaks this discrete symmetry softly.
The mass matrix becomes
\begin{eqnarray} M_6= \left(\begin{array}{cc}
  {\bf 0}                    &  \mu_d+ g_{dS}\vev{\phi_S} +ig_{dA}\vev{\phi_A}\\
  h'_d\vev{H^\dagger}{\bf 1} & \mu_D {\bf 1}
  \end{array}\right)
  \end{eqnarray}
To see that the one-loop correction to $\theta$ is zero in this
case, one can consider the one-loop corrections to each $3 \times 3$
block one by one.  For example, when the left upper block deviates
from ${\bf 0}$ and develop a small hermitian component, one can
show that the determinant maintains real.  This property follows
from the following lemma in the matrix algebra,
\begin{eqnarray}
\left | \begin{array}{cc}
      {\bf C} & {\bf A}  \\
      {\bf D} & {\bf B} \end{array} \right| = |  {\bf CB}-{\bf DA} |
\ , \hbox{ provided } {\bf CD}={\bf DC} \ .
\end{eqnarray}
Similarly when the other $3 \times 3$ block develops a small hermitian
correction, it is easy to show that the determinant remains.
Therefore the strong CP $\theta$ is zero at one-loop level.
\vskip.2cm \noindent{\bf SU(3) horizontal symmetry}\quad
Our second example uses $SU(3)$ horizontal symmetry to achieve the
same goal.  It is a modification of a model proposed by Masiero
and Yanagida\cite{Masiero:1998yi}.

The fermion spectrum is extended the same as the $SO(3)$ model
earlier.  The only difference is that each flavor is now transform
as a triplet under
a horizontal flavor symmetry $SU(3)_\parallel$, instead of
$SO(3)_\parallel$. In the scalar sector, there are new horizontal
neutral (inert to $SU(2)_L \times U(1)$) Higgs bosons, three
octets $\phi^a_\alpha$ ($a=1,\cdots,8$, $\alpha =1,2,3$). The
Yukawa couplings in the hamiltonian (for down quarks) are
\begin{eqnarray}
\bar d_R(g_{d\alpha}\phi^a_\alpha \lambda^a+m_D) D_L + \bar
D_R(g_{D\alpha}\phi^a_\alpha \lambda^a+M_D) D_L
                  \nonumber \\
+ h_d \bar D_R  H^\dagger q_L
+ h'_d \bar d_R  H^\dagger q_L
+ {\rm H.c.},
\end{eqnarray}
where $m_D$ and $M_D$ are bare masses, and similarly for up
quarks.
Three octets are needed in order that $SU(3)_\parallel$ is broken
completely and that both symmetric and antisymmetric components of
octet representation develop VEV.  CP is again assumed to be a
symmetry before any symmetry breaking and therefore all the Yukawa
couplings are assumed real.  As in previous model, $h_d$ (or
$h'_d$) can be set to zero by convention.  The mass matrix is then
\begin{eqnarray} (\bar d_R \  \bar D_R )
\left(\begin{array}{cc} {\bf 0} & m_D + g_{\alpha
d}\vev\phi_\alpha^a\lambda^a
\\h_d\vev{H^\dagger}{\bf 1} & M_D+g_{\alpha D}\vev\phi_\alpha^a\lambda^a
\end{array}\right) \left(\begin{array}{c} d_L \\ D_L
\end{array}\right) \ .
\end{eqnarray}
It is easy to see that the tree level strong CP $\theta$ vanishes.

To make one-loop $\theta$ vanishes, one can employ a discrete
symmetry as before.
Similar to Ref.\cite{Masiero:1998yi}, let $u, d$ and all $\phi$, $H$
fields be odd, and $U$ and $D$ even under the discrete symmetry.
So $g_{\alpha D}=0$, and the right lower block is $M_D{\bf 1}$.

Note that we need the soft term $m_D$ which breaks the discrete symmetry,
otherwise the hermitian mass
matrix will be traceless, not allowed phenomenologically. 
At a deeper lever, the $m_D$ term may come from the VEV of 
a $SU(3)_\parallel$ singlet $\Phi$ of odd parity under the discrete symmetry.
The $6\times 6$ matrix has real determinant when couplings are
real, as required by the imposed CP symmetry.
Note that to break CP and to have most general hermitian mass
matrix, one need all components of the octets to develop VEV, this
is why it is necessary for have three octets.  It is not clear
whether this has been proved in the literature or not.

Integrating out the exotic generation $D$, we have the effective
mass matrix for the known generations,
\begin{equation}
(1/M_D) h_d \vev{H^\dagger}
(m_D+ g_d\vev{\phi^a}\lambda^a) \ ,
\end{equation}
which is hermitian with real determinant. Some components of
Gell-Mann matrices $\lambda^a$ are complex to produce the desired
CKM phenomenology of CP violation.

It is interesting to note that the mass matrix is block-hermitian
because of spontaneous CP violation and the $SU(3)$ symmetry, the
fact that $\phi^\alpha$ are octet.  However, even after SU(3) is
broken, the block-hermiticity of the mass matrix maintains intact.

As long as the mass matrix is block-hermitian, the contribution to
$\theta$ will vanish to all order.  So, to look for contribution
to $\theta$, one looks for loop induced operator that may violate
this hermiticity, such as the higher dimensional operator,
$(1/M^n) \bar{d}_L d_R H f_n(\phi)$, where $f_n$ is a function of
$\phi$ or $\Phi$.  However, since the fermion bilinear can only
either be $SU(3)$ singlet or octet, $f_n$ has to be effective
singlet or octet. Since the effective octet will still give
hermitian form, the contribution to $\theta$ must come from the
effective singlet $f_n$.  Using the argument similar to that of
the case of $SO(3)$ discussed earlier, it is easy to see that in
this $SU(3)$ flavor model with discrete symmetry, the $\theta$ is
nonzero starting at the two-loop level.
\vskip.2cm \noindent{\bf Conclusion} \quad
We have shown that flavor symmetry properly arranged at high
energy can have the powerful consequence of producing a low energy
effective KM model with naturally small strong CP $\theta$.  We argue that the success of the KM model with very small $\theta$ may be providing us the clue of an flavor symmetry in high energy.  This letter provides two existence proofs of this idea, the next step is to tackle the harder question of actually using the flavor
symmetry to explain the fermion mass problem.

\vskip.2cm \noindent {\it Acknowledgment}: WYK is supported by the
U.S.~DOE under Grants No.~DE-FG02-84ER40173.  DC is supported by a
grant from NSC of Taiwan, ROC.


\begin{thebibliography}{99}
\bibitem{Kobayashi:fv}
M.~Kobayashi and T.~Maskawa,
Prog.\ Theor.\ Phys.\  {\bf 49} (1973) 652.

\bibitem{Shabalin:zz}
E.~P.~Shabalin,
Sov.\ J.\ Nucl.\ Phys.\  {\bf 36}, 575 (1982)
[Yad.\ Fiz.\  {\bf 36}, 981 (1982)].

\bibitem{Ellis:1978hq}
J.~R.~Ellis and M.~K.~Gaillard,
Nucl.\ Phys.\ B {\bf 150}, 141 (1979).

\bibitem{Khriplovich:1993pf}
I.~B.~Khriplovich and A.~I.~Vainshtein,
Nucl.\ Phys.\ B {\bf 414}, 27 (1994)
[arXiv:hep-ph/9308334].

\bibitem{Nelson:hg}
A.~E.~Nelson,
Phys.\ Lett.\ B {\bf 143}, 165 (1984);
Phys.\ Lett.\ B {\bf 136}, 387 (1984).
%
\bibitem{Barr:qx}
S.~M.~Barr,
Phys.\ Rev.\ Lett.\  {\bf 53}, 329 (1984),
Phys.\ Rev.\ D {\bf 30}, 1805 (1984).

\bibitem{Bento:ez}
L.~Bento, G.~C.~Branco and P.~A.~Parada,
Phys.\ Lett.\ B {\bf 267}, 95 (1991).

\bibitem{Glashow:2001yz}
S.~L.~Glashow,
arXiv:hep-ph/0110178.


\bibitem{ref:PQ}
R.~D.~Peccei and H.~R.~Quinn,
Phys.\ Rev.\ Lett.\  {\bf 38}, 1440 (1977);
R.~D.~Peccei and H.~R.~Quinn,
Phys.\ Rev.\ D {\bf 16}, 1791 (1977).

\bibitem{Li:1973mq}
L.~F.~Li,
Phys.\ Rev.\ D {\bf 9}, 1723 (1974).

\bibitem{Masiero:1998yi}
A.~Masiero and T.~Yanagida,
arXiv:hep-ph/9812225.




\end{thebibliography}
\end{document}